\documentclass[prl,aps,preprint,showpacs,showkeys]{revtex4}
\usepackage{amssymb}
\usepackage{graphicx}
\usepackage{amsmath}
\usepackage[usenames]{color}

\begin{document}

\title{ Magnetic quantum oscillations in the charge-density-wave state of
the organic metals $\alpha $-(BEDT-TTF)$_{2}$MHg(SCN)$_{4}$ with M = K and Tl%
}
\author{M. V. Kartsovnik$^{1}\footnote{Email: mark.kartsovnik@wmi.badw.de}$}
\author{V. N. Zverev$^{2,3}$}
\author{D. Andres$^1$\footnote{present address: Attocube Systems AG, 
K\"oniginstrasse 11a, 80539 M\"unchen, Germany}}
\author{W. Biberacher$^{1}$, T. Helm$^1$\footnote{present address: Material Science Devision,
Lawrence Berkeley National Laboratory 62R0203, 1 Cyclotron Road, 94720 Berkeley, CA, USA}}
\author{P. D. Grigoriev$^{4}$}
\author{R. Ramazashvili$^{5}$}
\author{N. D. Kushch$^{6}$}
\author{H. M\"uller$^{7}$}
\affiliation{$^{1}$Walther-Meissner-Institut, Bayerische Akademie der Wissenschaften,
Walther-Meissner-Strasse 8, D-85748 Garching, Germany}
\affiliation{$^{2}$Institute of Solid State Physics, Russian Academy of Sciences,
Academician Ossipyan Str. 2, Chernogolovka, 142432 Russia}
\affiliation{$^{3}$Moscow Institute of Physics and Technology, Dolgoprudny, Institutskii
9, Moscow reg., 141700, Russia}
\affiliation{$^{4}$ L. D. Landau Institute for Theoretical Physics, Russian Academy of
Sciences, Academician Semenov Ave. 1a, 142432 Chernogolovka, Russia}
\affiliation{$^{5}$Laboratoire de Physique Th\'eorique -- IRSAMC, CNRS and Universit\'e
de Toulouse, UPS, F-31062 Toulouse, France}
\affiliation{$^{6}$ Institute of Problems of Chemical Physics, Russian Academy of
Sciences, Academician Semenov Ave. 1, 142432 Chernogolovka, Russia}
\affiliation{$^{7}$European Synchrotron Radiation Facility, Rue Jules Horowitz 6, BP 220,
38043 Grenoble CEDEX 9, France}
\date{\today }

\begin{abstract}
The low-temperature charge-density-wave (CDW) state in the layered organic
metals $\alpha$--(BEDT-TTF)$_2$MHg(SCN)$_4$\ has been studied by means of
the Shubnikov -- de Haas and de Haas -- van Alphen effects. In addition to
the dominant $\alpha$-frequency, which is also observed in the normal state,
both the magnetoresistance and magnetic torque possess a slowly oscillating
component. These slow oscillations provide a firm evidence for the
CDW-induced reconstruction of the original cylindrical Fermi surface. The
$\alpha$-oscillations of the interlayer magnetoresistance exhibit an
anomalous phase inversion in the CDW state, whereas the de Haas -- van
Alphen signal maintains the normal phase. We argue that the anomaly may be
attributed to the magnetic-breakdown origin of the $\alpha$-oscillations in
the CDW state. A theoretical model illustrating the possibility of a phase
inversion in the oscillating interlayer conductivity in the presence of a
spatially fluctuating magnetic breakdown gap is proposed.
\end{abstract}

\maketitle

\section{Introduction}

Comprehensive quantitative description of the de Haas -- van Alphen effect
by the Lifshitz-Kosevich theory has made magnetic quantum oscillations (MQO)
one of the most powerful tools for studying conduction electrons in metals.
This tool has been extensively used not only for exploring conventional
metals \cite{crac73} but also for gaining a deep insight into electronic
systems of more complex materials such as cuprate \cite{tail09,kart11} and
iron-based \cite{walm13,graf12} high-temperature superconductors, heavy
fermion compounds \cite{tail87}, and organic charge-transfer salts \cite%
{kart08a}. Particularly the latter class of materials has demonstrated the
great potential of the de Haas -- van Alphen (dHvA) and Shubnikov -- de Haas
(SdH) effects in revealing the Fermi surface properties in various
electronic states. Additionally, the organic compounds, generally
characterized by an extraordinary crystal quality, very high anisotropy, and
significant electron interactions, offer a vast playground for studying
specific features of MQO in layered correlated electron systems, see for a
review Refs. \cite{kart04,kart08a,kart05,grig13} and references therein.

A spectacular example of how MQO and high-field classical magnetoresistance
can be used for investigating the electronic state of an organic metal is
the work done on $\alpha$--(BEDT-TTF)$_2$MHg(XCN)$_4$, where BEDT-TTF stands
for the donor organic molecule
bis\-(ethylene\-dithio)\-tetra\-thia\-fulvalene, M\,= K, Tl, NH$_4$, and Rb,
and X\,= S, Se, see \cite{kart08a} for a review. These are isostructural
layered charge-transfer salts with the Fermi surface comprising a pair of
slightly warped open sheets
(representing a quasi-one-dimensional, q1D, conduction band)
and a cylinder (a quasi-two-dimensional, q2D, band) \cite{mori90a,rous96}.
The compounds display a huge electronic anisotropy:
the ratio of the effective transfer integrals within and across conducting
layers $t_{||}/t_{\perp}$ is in the range $10^2-10^3$ \cite{hana01,kart06}.
As a result, the MQO have a very large amplitude and bear a pronounced 2D
character \cite{harr96,lauk95,hono98,wosn00,grig12}. The monotonic part of
the interlayer magnetoresistance also shows severe deviations from the
conventional three-dimensional behavior \cite{kart06,grig12,kart09b}.

At temperature $T \sim 10$\,K three salts, with M\,= K, Tl, and Rb, and X\,=
S undergo a charge-density-wave (CDW) transition caused by the Peierls-type
nesting instability of the open Fermi sheets \cite{kart08a,four10}. The
compounds remain, however, metallic due to the ungapped cylindrical Fermi
surface. The low-temperature state is characterized by a bunch of striking
anomalies in high magnetic fields which, actually, have triggered the
initial interest in these materials \cite%
{osad90,sasa90a,kart92b,kart93,caul95}. By now, it is clear that their
behavior is largely governed by the coexistence of a narrow-gap CDW and
metallic q2D carriers and, consequently, a rich phase diagram including
several kinds of magnetic field-induced transitions between different CDW
states, see, e.g., \cite{kart97,harr00,andr03,andr11a}. However, a number of
anomalies are still a matter of debate. One of the problems in this respect
is that there is no general consensus as to the exact topology of the
reconstructed Fermi surface in the CDW state \cite%
{kart93,harr99a,harr04,kang13}. Moreover, even the occurrence of
reconstruction itself has been questioned \cite{yosh95a,maki03}.

Here we report on experimental studies of magnetic quantum oscillations in
the CDW state of $\alpha$--(BEDT-TTF)$_2$MHg(SCN)$_4$\ with M\,= K and Tl.
In addition to the dominant frequency corresponding to the large cylindrical
Fermi surface predicted by the normal-state band structure calculations
\cite{mori90a,rous96}, a new, low frequency is found in both the SdH and
dHvA spectra. This result is discussed in terms of the Fermi surface
reconstruction in the CDW state. Further, the phase of the SdH (but not
dHvA!) oscillations in the low-temperature, low-field CDW state is shown to
be inverted in comparison to that in the normal state. We propose a model
that qualitatively explains this anomalous behavior via spatial fluctuations
of the magnetic breakdown gap.

\section{Experimental}

The experiments were carried out on single crystals of
$\alpha$--(BEDT-TTF)$_2$KHg(SCN)$_4$\ and
$\alpha$--(BEDT-TTF)$_2$TlHg(SCN)$_4$\ hereafter
referred to as the K-salt and Tl-salt, respectively. The samples were
submillimeter-size platelets of a distorted hexagon shape with large faces
parallel to the highly conducting BEDT-TTF layers. For ambient-pressure
studies a setup allowing simultaneous measurements of the interlayer
magnetoresistance and magnetic torque \cite{weis99b} was used. High-pressure
magnetoresistance experiments were done using a small clamp pressure cell
made of nonmagnetic Cu-Be alloy. The pressure was applied at room
temperature and its low-$T$ value was determined from the resistance of a
calibrated manganin pressure gauge. The oscillations of magnetoresistance
and magnetic torque were studied in the temperature range from 0.4 to
4.2\,K. Magnetic fields up to 17\,T were generated by a superconducting
solenoid. Experiments at higher fields, up to 29\,T were conducted at the
Laboratoire National des Champs Magn\'{e}tiques Intenses (LNCMI), Grenoble,
France.

\section{MQO spectrum in the CDW$_0$ state}

\begin{figure}[tb]
\includegraphics[width=8cm]{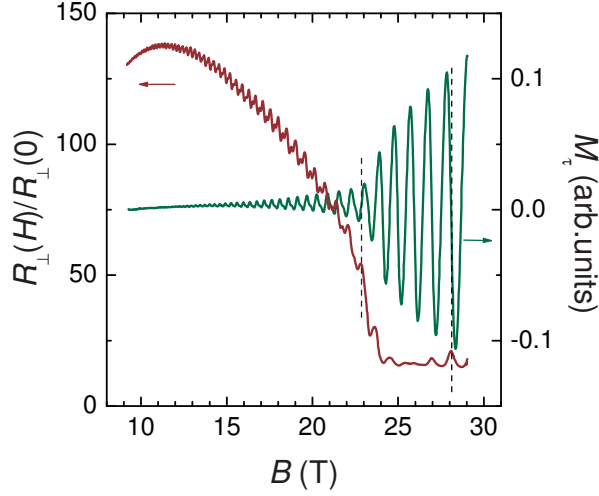}
\caption{Field-dependent interlayer resistance (left-hand scale)
and magnetization (right-hand scale) of the K-salt in the CDW state;
$T=1.4$\,K; the field is tilted by angle $\theta =6.2^{\circ}$ from the
direction perpendicular to the conducting layers.}
\label{restorq}
\end{figure}

The magnetoresistance and magnetization of the
$\alpha$--(BEDT-TTF)$_2$MHg(SCN)$_4$\ salts in the CDW state were
studied by many authors, see \cite{kart08a} and references therein.
The general behavior in a field nearly perpendicular to conducting
BEDT-TTF layers is illustrated in Fig.~\ref{restorq}. Below the
so-called kink field, $\mu_0H_k=24$ and 27\,T for the
K- and Tl-salts, respectively \cite{osad90,kart94}, the zero-field CDW$_0$
state is stable. As seen from Fig.\,\ref{restorq}, it is characterized by a
very high interlayer magnetoresistance showing a peak at around 10\,T and
then gradually turning down. The MQO in the CDW$_0$ state are moderately
strong, of the order of a few percent of the total signal, and have a
distorted or even split shape due to an anomalously strong second harmonic
contribution. At $\mu_0H_k$ the system is driven into the CDW$_x$ state
\cite{chri00,prou00} with a $B$-dependent spatially modulated order parameter
analog of the Larkin-Ovchinnikov-Fulde-Ferrel state predicted for
superconductors \cite{buzd83c,zanc96,grig05}. Both the dHvA and SdH
oscillations are strongly enhanced upon entering this state: for example,
the SdH amplitude in Fig.\,\ref{restorq} amounts to $\approx 30\%$ of the
nonoscillating background at 28\,T.

\begin{figure}[tb]
\includegraphics[width=8cm]{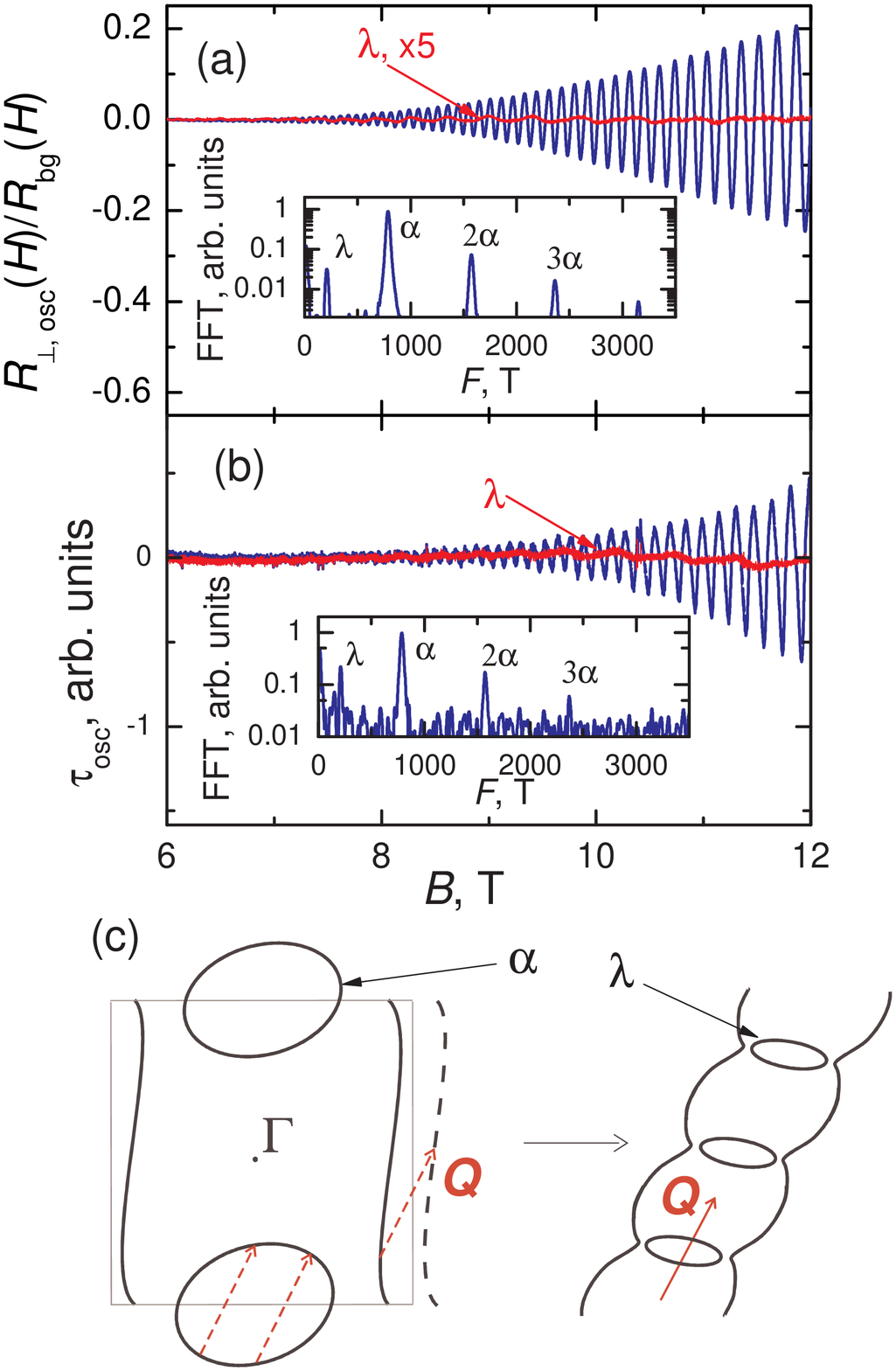}
\caption{Oscillating components of the
magnetoresistance (a) and magnetic torque (b) of the K-salt at $T=0.45$\,K,
$\theta=31.5^{\circ}$. The red curves are obtained by filtering out the
$\alpha$-oscillations and demonstrate the behavior of the slow oscillations
with frequency $F_{\lambda}=210$\,T. In (a) the $\lambda$-oscillations are
magnified by a factor of 5, for a better visibility. The insets in (a) and
(b) show the corresponding fast Fourier spectra. (c) Schematic 2D view of
the Fermi surface reconstruction due to the CDW potential with the wave
vector $Q$. The original Fermi surface (left panel) consists of a pair of
open sheets and a cylinder. The CDW, introducing a new periodicity with the
wave vector $\mathbf{Q}$, opens a gap at the Fermi level in the whole open
branch as well as in the q2D band at the states separated by $\mathbf{Q}$
(right panel).}
\label{lambda}
\end{figure}

Fig. \ref{lambda} shows the oscillatory components of the interlayer
magnetoresistance (a) and torque (b) of the K-salt in the CDW$_0$ state and
(inset) their fast Fourier transform (FFT) spectra. In order to obtain an
appreciable signal in the torque, the field is tilted by the angle
$\theta=31.5^{\circ}$ from the normal to the layers. The oscillations are
dominated by the fundamental frequency $F_{\alpha}=787$\,T. In agreement
with earlier experiments, this frequency, recalculated to the purely
out-of-plane field orientation, $F_{\alpha,0}=F_{\alpha}(\theta)\cos\theta
=670$\,T, corresponds to the cylindrical Fermi surface with the same area as
in the normal state, see Fig. \ref{lambda}(c).

In addition to $F_{\alpha}$ and its higher harmonics, a low frequency peak
at $F_{\lambda}=210$\,T is clearly pronounced in the FFT spectra. It is
important that the low frequency is observed not only in magnetoresistance
but also in magnetization. The latter establishes its thermodynamic origin,
ruling out kinetic effects like quantum interference of open electron
trajectories \cite{star77} or due to a weak warping of the Fermi cylinder
\cite{grig03} which are often found in layered organic metals \cite%
{kart04,kart05}. We, therefore, attribute it to a real closed orbit on a
small cylindrical Fermi surface undergoing Landau quantization in a strong
magnetic field.

The present result provides a solid argument in favor of the Fermi surface
reconstruction model based on studies of the semiclassical angle-dependent
magnetoresistance oscillations (AMRO) \cite{kart93}. According to this
model, the CDW potential $V_{\mathbf{Q}}$ with the wave vector $\mathbf{Q}$,
besides nesting the open sheets of the original Fermi surface, sketched in
the left panel of Fig.~\ref{lambda}(c), also has an effect on the q2D band.
It folds the cylindrical part of the Fermi surface, mixing the states with
the wave vectors $\mathbf{k}$ and $\mathbf{k}+\mathbf{Q}$. This creates a
new small cylinder and a pair of strongly corrugated sheets extending along
$\mathbf{Q}$, as shown in the right panel of Fig.~\ref{lambda}(c). The open
sheets are responsible for the AMRO with sharp dips at the so-called Lebed
magic angles \cite{kart93,kart08a,kang13}. The small cylindrical Fermi
surface is manifested through the low-frequency MQO, $F_{\lambda}$. The
$\alpha$-oscillations come as a result of magnetic breakdown (MB).

The energy scale of the gap $\Delta$ separating the open and closed parts of
the Fermi surface is of the order of the CDW potential, which can be
assessed from the critical field of the low-temperature CDW$_0$-CDW$_x$
phase transition, $\Delta \sim \pi\mu_{\mathrm{B}}B_k$ \cite%
{buzd83c,zanc96,grig05}. For the K-salt, $B_k \approx 24$\,T, yielding
$\Delta_{\mathrm{(}CDW)} \sim 4$\,meV. Then, using the Blount criterion for
MB \cite{shoe84}, one can obtain an estimate of the breakdown field,
$B_{\mathrm{MB}} \sim \Delta^2m_c/\hbar eE_{\mathrm{F}} \sim 6$\,T, where
$E_{\mathrm{F}} \simeq 35$\,meV and $m_c \approx 1.5m_e$ are, respectively,
the Fermi energy and effective cyclotron mass evaluated from the MQO data
and $e$ is the elementary charge.

At fields $\sim 8$\,T, at which the MQO just become resolvable, the
amplitudes of the $\lambda$- and $\alpha$-oscillations coming from the
classical and magnetic-breakdown orbits, respectively, are comparable. As
the field increases further, the breakdown probability grows exponentially.
This leads to a rapid enhancement of the $\alpha$-oscillations by contrast
to the almost constant amplitude of the $\lambda$-oscillations, c.f. blue
and red lines in Fig.~\ref{lambda}(a),(b). Above 17\,T the relative
contribution of the $\lambda$-frequency to the MQO spectrum becomes
vanishingly small due to the strong magnetic breakdown. The latter is also
reflected in the negative slope of the magnetoresistance, see
Fig.~\ref{restorq}, as well as in the behavior of the AMRO \cite{hous96b}.
As the system enters the CDW$_x$ state at $B_k$, the CDW gap is
considerably reduced \cite{zanc96,grig05}. The probability of magnetic
breakdown increases to almost unity; both the MQO and classical
magnetoresistance are fully determined by the breakdown orbit $\alpha$,
as if the cylindrical Fermi surface were unreconstructed. In particular,
this is a reason why the high-field CDW$_x$ state was for some time
confused with a reentrant normal state \cite{sasa92,hous96d}.

Besides the frequencies $F_{\alpha}$ and $F_{\lambda}$, their linear
combinations have been observed in some experiments \cite%
{kart93,caul95,hous96b}, which is consistent with the proposed model of the
Fermi surface reconstruction. An additional frequency of 775\,T has been
reported once for the K-salt that would be difficult to account for.
However, to the best of our knowledge, it has not been reproduced by other
authors. It should be noted that there is still no simple explanation of the
anomalously strong second harmonic of $F_{\alpha}$ observed in the CDW$_0$
state at field orientations nearly normal to the layers. In spite of several
attempts to explain them \cite{sasa99,harr01}, this puzzle is still awaiting
a convincing solution.

\section{Phase inversion of the SdH oscillations}

There is a notable difference in the behavior of the dHvA and SdH signals in
Fig.~\ref{restorq}. While the magnetization displays regular oscillations
continuously growing with field, the SdH signal has a node-like feature at
26~T, i.e. shortly after entering the CDW$_{x}$ state. The phase of the SdH
oscillations inverts at the node. On the high-field side the positions of
the resistance peaks coincide with the midpoints of the decaying
half-periods of the magnetization oscillations corresponding to the integer
filling factors (the chemical potential resides in the middle between the
adjacent Landau levels). This is exactly what is expected of the interlayer
magnetoresistance oscillations in the bulk q2D regime, when the Landau level
(LL) spacing $\hbar \omega _{c}$ exceeds the interlayer bandwidth $4t_{\perp
}$ \cite{harr96,cham02b,grig13,grig11}. At first glance, the inverted phase
of the SdH oscillations below the node-like feature might be attributed to a
dimensional crossover from the high-field q2D regime to a more conventional
3D one, where the resistance should peak at odd-half-integer filling
\cite{shoe84,pipp65}. This, however, is hardly the case: due to a very weak
interlayer coupling, the dimensional crossover in the present materials
takes place already at 1--2~T \cite{kart06,grig12}. Therefore, the
oscillation phase below the node should be considered as anomalous.

\begin{figure}[tb]
\includegraphics[width=8cm]{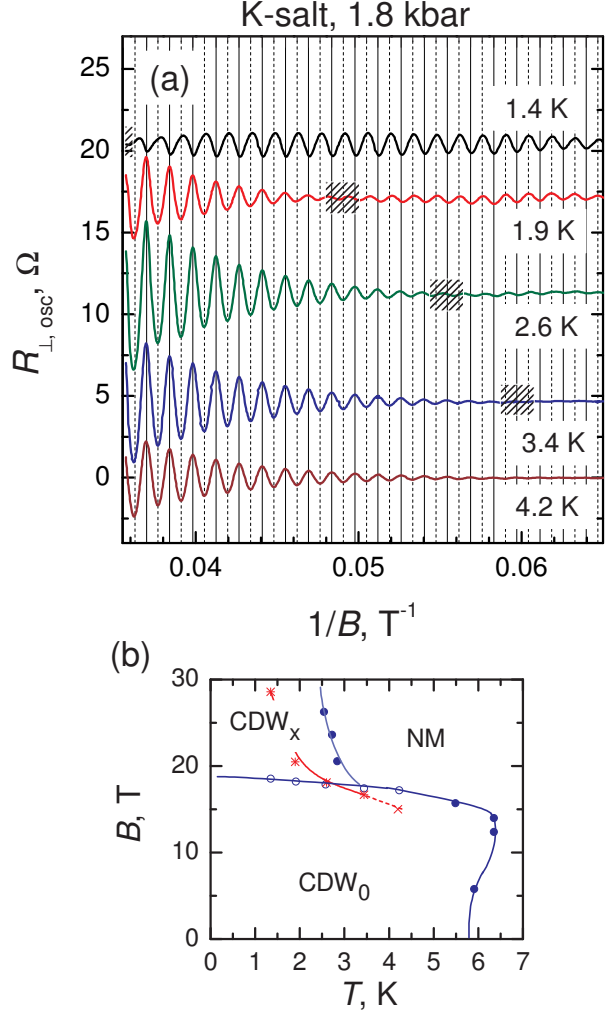}
\caption{Oscillations of the interlayer resistance
of the K-salt under pressure $P = 1.8$\,kbar plotted in the $1/B$ scale, for
different temperatures. The vertical solid (dashed) grid lines correspond to
the integer (odd half-integer) filling factors, see text. The hatched boxes
indicate the regions where the oscillation phase changes by $\pi$.
(b) The $T-B$ phase diagram (filled and empty
blue circles), taken from Ref.~\cite{andr01}, and the phase inversion points
(red stars) determined from the data in (a).}
\label{KHgpress}
\end{figure} 
At ambient pressure, the phase inversion occurs only in the CDW$_x$ state,
above 24\,T that makes it rather difficult to trace in steady magnetic
fields. Therefore, we have performed magnetoresistance measurements under
pressure which is known to shift the CDW phase boundaries to lower fields
and temperatures \cite{andr01}. The results of the experiment on the K-salt
under a pressure of 1.8\,kbar, in the field perpendicular to the layers are
summarized in Fig.~\ref{KHgpress}. In panel (a) the SdH oscillations,
obtained by subtracting a low-order polynomial from the as-measured
interlayer resistance, are plotted as a function of inverse field for
different temperatures. The solid (dashed) grid lines are drawn through the
maxima (minima) of the resistance at 4.2\,K. The lower-temperature curves
show an inversion of the oscillation phase within narrow field intervals
marked by hatched boxes. Following the results of the ambient pressure
experiment, we conclude that also under pressure the SdH phase is "correct"
(resistance maxima at integer filling factors) at high fields and anomalous
at lower fields.

Note that the field $B_{\mathrm{pi}}$, at which the phase inversion occurs,
rapidly increases at lowering the temperature. This obviously is reflected
in the temperature dependence of the oscillation amplitude, leading to
drastic deviations from the conventional Lifshitz-Kosevich behavior
\cite{hill97a,hono00}. Indeed, according to the data in Fig.~\ref{KHgpress}(a),
the $T$-dependence of the SdH amplitude measured at around 28\,T ($%
1/B=0.036\;\mathrm{T}^{-1}$) reaches its maximum at 2.6\,K and nearly
vanishes at 1.4\,K; at fields 25-26\,T the amplitude has a minimum near
2\,K. Therefore, one has to be extremely careful when trying to analyze the
$T$-dependence of the SdH amplitude in the present system in terms of the
Lifshitz-Kosevich theory.

The temperature-dependent field $B_{\mathrm{pi}}$ is plotted in
Fig.~\ref{KHgpress}(b) (stars) on top of the phase diagram of the K-salt
at 1.8\,kbar (circles). For $T=4.2$\,K no phase inversion was detected
down to 15\,T, the lowest field at which the oscillations could still be
resolved; therefore, the point at 4.2\,K only indicates the upper limit for
$B_{\mathrm{pi}}$. From this plot it becomes clear that the anomalous SdH
phase exists only in the CDW state, at temperatures and fields sufficiently
distant from the boundary to the normal state.

\begin{figure}[tb]
\includegraphics[width=8cm]{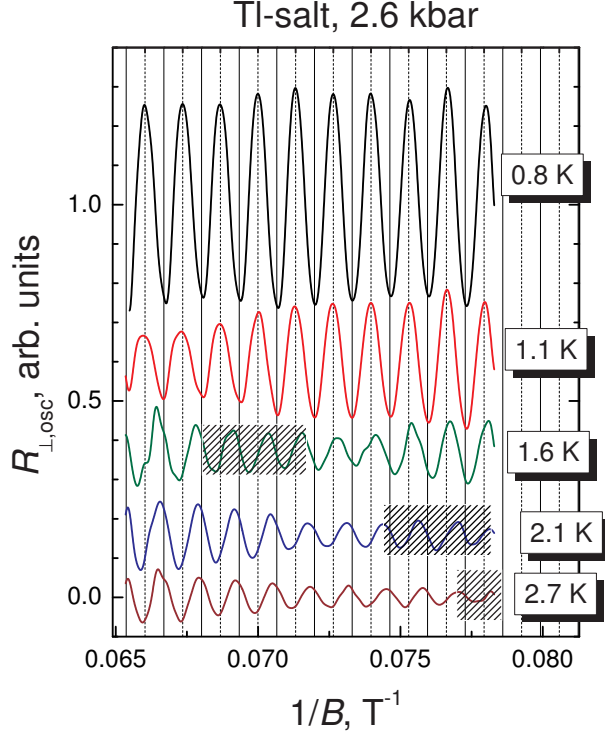}
\caption{The same plot as in Fig. 3(a) but for the
Tl-salt under pressure $P=2.6$\,kbar.}
\label{TlHgpress}
\end{figure}

To verify that the anomalous SdH phase is a general property of the CDW
state of $\alpha$--(BEDT-TTF)$_2$MHg(SCN)$_4$\ and not just a feature of the
K-salt, we made similar measurements on the Tl-salt under a pressure of
2.6\,kbar. Fig.~\ref{TlHgpress} shows the oscillating component of
magnetoresistance in the inverse-field scale. While no perfect nodes are
observed in this case, the oscillations clearly invert their phase upon
changing field and temperature. Note that the data in Fig.~\ref{TlHgpress}
correspond to the field range below 15.5\,T which is significantly below the
CDW$_0$ -- CDW$_x$ transition. Thus, the phase inversion in the field sweeps
occurs deep inside the CDW$_0$ phase for temperatures above 1.5\,K.

The inversion of the SdH signal in the $\alpha$--(BEDT-TTF)$_2$MHg(SCN)$_4$\
compounds has already been reported by a number of authors \cite%
{hill97a,hono00,harr00,andr01a,uji08}. Most of these experiments were
focused on the high-field state, above the kink field $B_k$. Consequently,
all the proposed explanations, involving various exotic phenomena such as
bulk quantum Hall effect \cite{hill97,hono00}, Froehlich superconductivity
\cite{harr00}, or an unconventional quantum liquid \cite{harr04}, associated
the anomalous phase with the high-field CDW$_x$ or even normal state. They
all implied a normal behavior to be restored at lower fields as the system
enters the CDW$_0$ state and/or the oscillations become weak. This obviously
contradicts the present results showing that the anomalous phase does exist
in the CDW$_0$ state. Moreover, no re-entrance to the normal phase is
observed at lowering the field as long as the oscillations can be resolved.
Therefore, the mechanism responsible for the phase inversion should be based
on a property common for the CDW$_0$ and CDW$_x$ states. A possible
candidate is magnetic breakdown between the open and closed parts of the q2D
Fermi surface. Below we propose a model, illustrating how magnetic breakdown
may lead to the phase inversion of the SdH signal in a system with spatial
fluctuations of the breakdown gap, and argue that it is qualitatively
consistent with the present experiment.

\section{Model of spatially inhomogeneous magnetic breakdown}

The observed magnetoresistance is due to two subsets of electron
orbits in the momentum space.
The first one appears due to MB and corresponds to the closed
electron trajectory along the $\alpha $-pockets of the Fermi surface, see
Fig.~\ref{lambda}(c). This orbit is responsible for the quantum oscillations with the
frequency $F_{\alpha }$ and would be the only one observed for the breakdown
amplitude $p=1$. The second subset includes all other trajectories on the
reconstructed Fermi surface: open trajectories on the strongly corrugated
open sheets, closed orbits on the very small FS cylinders, yielding the MQO
frequency $F_{\lambda }$, as well as a variety of other MB orbits, which
do not contribute to the $\alpha $-frequency of MQO. The main contribution
to the electron density of states (DoS) of the second subset comes from the
states on the open non-quantized orbits. Hereafter we refer to this second
subset as to the open or q1D branch, whereas the first subset, producing
the $\alpha $-frequency, we call the q2D branch.
At a finite MB amplitude both branches are present.
To encircle the $\alpha $-orbit, electrons must undergo four
MB transitions, see Fig.~\ref{lambda}(c). Therefore, the amplitude to complete
the $\alpha $-orbit is $p^{4}R_{D}$ \cite{fali66}, where $p$ is MB amplitude
and $R_{D}$ is the Dingle factor coming from the electron scattering on
short-range defects \cite{CommentRD}; in clean samples $R_{D}$ is close to
unity. Hence, in analogy to the case of MQO in the regime of a finite MB
amplitude $p$ \cite{fali66}, we may very roughly evaluate the density of  q2D
electron states as $\rho _{2D}\left( \varepsilon \right) \approx |p|^{4}\rho
\left( \varepsilon \right) $, and the density of the q1D states as $\rho
_{1D}\left( \varepsilon \right) =(1-|p|^{4})\rho \left( \varepsilon \right)$,
where $\rho \left( \varepsilon \right) $ is the total DoS \cite{CommentQMB}.
In our experiment the MB amplitude at $B=20$~T is close to unity:
$|p|=\exp (-B_{\mathrm{MB}}/2B)\simeq 0.86$, using the given above estimate
$B_{\mathrm{MB}}\simeq 6$\thinspace T. However, the relative weight of the
q1D states is still comparable to that of  q2D  states: $1-p^{4}\approx 0.45$.
The significant contribution from the q1D states is manifested, for instance,
in the AMRO behavior.

The MB with a finite amplitude induces transitions between these two
branches: from the q1D to the  q2D  branch with the amplitude $p$ and probability
$|p|^{2}$, and to the q1D branch with the amplitude $q=e^{i\phi }\sqrt{1-|p|^{2}}$
and probability $1-|p|^{2}$. For an ideal crystal, the MB amplitude does not
depend on coordinates. However, in various organic metals the CDW order
parameter is subject to spatial fluctuations, as in the
soliton phase of the CDW (see, e.g., Refs. \cite{braz84} and \cite{su81} for
review). These solitons locally reduce the CDW gap value $\Delta _{\mathrm{CDW}}$.
Since the CDW gap defines the MB gap, namely, $B_{\mathrm{MB}}\propto \Delta
_{\mathrm{CDW}}^{2}$, the MB amplitude $p$ significantly increases in such soliton
spots. For simplicity, consider the MB amplitude $p=p_{0}\left( B\right) $
everywhere except for certain "MB defect" spots, where $p=p_{1}\approx 1$.
This means that the defect spots scatter the electrons to the  q2D  states.
The scattered electrons change their momentum in the direction perpendicular
to the layers ($z$-direction) because the MB defects are local, which leads
to relaxation of the $z$-component of electron momentum as if due to impurities.

In spite of an obvious similarity between the MB-defect spots and randomly
distributed impurities, there is an important difference between the two:
the latter scatter electrons to any state with the same energy, and in the
Born approximation the corresponding scattering rate is $1/\tau
_{i}\propto \rho \left( \varepsilon \right) $, where the total DoS $\rho
\left( \varepsilon \right)$ is the sum of the q1D and the q2D  DoS:
$\rho \left(\varepsilon \right) =\rho _{1D}\left( \varepsilon \right) +
\rho _{2D}\left(\varepsilon \right)$. In our model, the MB defects
scatter  electrons only to the  q2D  branch, and their scattering rate
\begin{equation}
1/\tau _{\mathrm{MB}}\propto \rho _{2D}\left( E_{\mathrm{F}}\right) .
\label{tauMB}
\end{equation}

In the $\tau $-approximation the interlayer conductivity
\begin{equation}
\sigma _{zz}=2e^{2}\tau _{\mathrm{tot}}\sum_{\boldsymbol{k}, \alpha}\,v_{z\, \alpha}^{2}
\left( \boldsymbol{k}\right) \left( -n_{F}^{\prime }\left[ \varepsilon _\alpha
\left(\boldsymbol{k}\right) \right] \right) ,  \label{s1D}
\end{equation}
where $\alpha$ labels the branch, the total scattering rate
is given by the sum of the contributions from MB defects and from impurities:
\begin{equation*}
1/\tau _{\mathrm{tot}}=1/\tau _{\mathrm{MB}}+1/\tau _{i},
\end{equation*}
and $\left\langle v_{z\,1D(2D)}^{2}\right\rangle $ is the average square of
electron velocity component $v_{z}$ on the Fermi level of the q1D (q2D)
part of the spectrum. The derivative of the Fermi distribution function is
$n_{\mathrm{F}}^{\prime }(\varepsilon )=
-1/\{4T\cosh ^{2}\left[ \left( \varepsilon -E_{
\mathrm{F}}\right) /2T\right] \}\rightarrow \delta \left( \varepsilon -E_{
\mathrm{F}}\right) $ at $T\rightarrow 0$, and Eq. (\ref{s1D}) simplifies to
\begin{equation*}
\sigma _{zz}=2e^{2}\tau _{\mathrm{tot}}\left( E_{\mathrm{F}}\right) \left\langle
v_{z\,}^{2}\right\rangle \rho \left( E_{F}\right) .
\end{equation*}
\ In very clean samples, $1/\tau _{\mathrm{MB}}\gg 1/\tau _{i}$, so that the
main relaxation of electron momentum arises from the scattering off MB
defects:
\begin{equation}
1/\tau _{\mathrm{tot}}\approx 1/\tau _{\mathrm{MB}}.  \label{tt}
\end{equation}
Incidentally, the increasing scattering rate in the MB regime may account
for the unusually strong magnetoresistance of the present compounds at
$B\sim 10$\thinspace T, see Fig.~\ref{restorq}.

Combining Eqs. (\ref{tauMB})-(\ref{tt}) we obtain
\begin{equation}
\sigma _{zz}\propto \frac{\left\langle v_{z1D\,}^{2}\right\rangle \rho
_{1D}\left( E_{\mathrm{F}}\right) +\left\langle v_{z2D\,}^{2}\right\rangle
\rho _{2D}\left( E_{\mathrm{F}}\right) }{\rho _{2D}\left( E_{\mathrm{F}
}\right) }.  \label{sz}
\end{equation}
Due to the Landau quantization, the  q2D  DoS $\rho _{2D}\left( E_{\mathrm{F}
}\right) $ is an oscillating function of $E_{\mathrm{F}}/\hbar \omega _{c}$
around some constant value $\rho _{2D0}$. For the q1D branch,
there is no LL quantization, and the q1D DoS does not oscillate:
$\rho_{1D}\left( E_{\mathrm{F}}\right) \approx \rho _{1D0}$.

The mean-square velocity on the  q2D  branch $\left\langle
v_{z\,2D}^{2}\right\rangle $ is also an oscillating function of
$E_{\mathrm{F}}/\hbar \omega _{c}$, but the amplitude and even the sign of its
oscillations depends on the ratio of $t_\perp$, $\hbar \omega _{c}$ and $\hbar /\tau$.
In 3D metals with $t_\perp \gg \hbar \omega _{c}$ the oscillations of $\left\langle
v_{z2D\,}^{2}\right\rangle $ are weak and in the opposite phase to the DoS
oscillations, which determines the phase of SdH oscillations in 3D
metals. In almost 2D metals with $t_\perp \ll \hbar \omega _{c}$ but for weak
MQO, the oscillations of $\left\langle v_{z2D\,}^{2}\right\rangle $ are much
weaker than the oscillations of $\rho _{2D}\left( E_{\mathrm{F}}\right) $.
To show this, consider $\left\langle v_{z\,2D}^{2}\right\rangle =I\left( E_{
\mathrm{F}}\right) /\rho _{2D}\left( E_{\mathrm{F}}\right) $, where the
quantum oscillations of $I\left( E_{\mathrm{F}}\right) \equiv
\sum_{FS}v_{z\,2D}^{2}$ are given by Eq.(3) of Ref. \cite{kart02a} and the
oscillations of $\rho _{2D}\left( E_{\mathrm{F}}\right) =1/\tau _{i}\left(
E_{\mathrm{F}}\right) $ are given by the Eq. (2) of Ref. \cite{kart02a}.
At $t_\perp \ll \hbar \omega _{c}$, if we keep only the fundamental
harmonic of MQO, this gives
\begin{equation}
\rho _{2D}(\varepsilon )\propto 1-2R_{D}\cos \left( 2\pi \varepsilon /\hbar \omega
_{c}\right) ,  \label{DoS}
\end{equation}
where $R_{D}$ is the Dingle factor, and $I(\varepsilon )\propto 1-2R_{D}\cos
\left( 2\pi \varepsilon /\hbar \omega _{c}\right) \propto \rho _{2D}(\varepsilon )$
, i.e. $\left\langle v_{z\,2D}^{2}\right\rangle =I\left( E_{\mathrm{F}
}\right) /\rho _{2D}\left( E_{\mathrm{F}}\right) =\mathrm{const}$ up to the
terms of the order of $R_{D}^{2}$. Hence, for $R_{D}\ll 1$ at $t_\perp \ll
\hbar \omega _{c}$,
\begin{equation}
\left\langle v_{z\,2D}^{2}\right\rangle \approx \left\langle
v_{z\,1D}^{2}\right\rangle \approx 2t_{\perp }^{2}d^{2}/\hbar ^{2}.
\label{vz}
\end{equation}
For strong MQO in almost 2D metals, i.e. when $R_{D}\approx 1$
and $t_\perp \ll \hbar \omega _{c}$, the oscillations of $\left\langle
v_{z\,2D}^{2}\right\rangle $ cannot be neglected and must be calculated
beyond the $\tau $-approximation \cite{cham02b,grig11,grig13}. In this
limit the oscillations of $\left\langle v_{z\,2D}^{2}\right\rangle $ are
in phase with the oscillations of $\rho _{2D}(\varepsilon)$:
\begin{equation}
\left\langle v_{z\,2D}^{2}\right\rangle (\varepsilon )\approx \left\langle
v_{z\,1D}^{2}\right\rangle \left[ 1-2\beta R_{D}\cos \left( 2\pi \varepsilon
/\hbar \omega _{c}\right) \right] ,  \label{vzo}
\end{equation}
where the effective parameter $\beta \lesssim 1$ depends on magnetic field.

Substituting Eqs. (\ref{DoS}) and (\ref{vzo}) to Eq. (\ref{sz}) we obtain
\begin{equation}
\sigma _{zz}\propto \mathrm{const}+\left( \frac{\rho _{1D0}}{\rho _{2D0}}-\beta
\right) 2R_{D}\cos \left( 2\pi F/B\right) .  \label{szz1}
\end{equation}
For comparison, in the case of impurity scattering only, i.e.
without MB defects, in a quasi-2D limit we have\cite{shoe84}
\begin{equation}
\sigma _{zz}\propto 1-2R_{D}\cos \left( 2\pi F/B\right) .  \label{szz0}
\end{equation}
Hence, within the model above, the SdH oscillations change their sign
when the ratio $\rho _{1D0}/\rho _{2D0}$ crosses $\beta $. Thus, spatial
fluctuations of magnetic breakdown may lead to a phase inversion of the SdH
signal in a field sweep. Since the MB gap, determined by the CDW potential
is small near the CDW -- normal-metal boundary, the anomalous phase only
occurs at certain temperatures well enough below the CDW transition. This
explains the $T$-dependence of the phase inversion field. Finally, according
to Fig.~\ref{KHgpress}(b), the anomalous phase is apparently less stable in
the CDW$_{x}$ state, which has a much larger concentration of solitons.
Within the proposed model, this is obviously related to the significant drop
of the CDW gap upon entering this high-field state.

\section{Acknowledgements}

This work has been supported by the German Research Foundation
grant KA 1652/4-1, by the Russian Foundation for Basic Research grants Nos.
12-02-00312 and 13-02-00178, and by SIMTECH Program grant 246937. The
experiments in fields above 15\thinspace T were supported by the LNCMI-CNRS,
member of the European Magnetic Field Laboratory (EMFL). The visit of P.
D. G. to the LPT Toulouse was supported by l'Agence Nationale de la Recherche
under the program ANR-11-IDEX-0002-02, reference ANR-10-LABX-0037-NEXT.

\newpage 

\begin{thebibliography}{99}
\bibitem[Com(RD)]{CommentRD}
\bibinfo{note}{The total Dingle factor, measured
from MQO, has contribution not only from the electron scattering
on short-range defects but also from the long-range smooth
potential, giving the local shift of the Fermi energy (see Refs.
\cite{kart02a,grig12} for details).}
\bibitem[Com()]{CommentQMB}
\bibinfo{note}{The MB amplitude is also energy-dependent: $p=p\left(
  \varepsilon \right) $. Below we take the values $\rho \left( \varepsilon
  \right) ,\rho _{1D}\left( \varepsilon \right) ,\rho _{2D}\left( \varepsilon
  \right) $ and $p\left( \varepsilon \right) $ at the Fermi level.}
\end{thebibliography}

\end{document}